\documentclass[10pt,letterpaper,compsoc,conference]{iiswc26}

\usepackage{cite}
\usepackage{amsmath,amssymb,amsfonts}
\usepackage{algorithmic}
\usepackage{graphicx}
\usepackage[dvipsnames]{xcolor}
\usepackage[final]{microtype}
\usepackage[italic]{mathastext}
\usepackage{libertine}
\usepackage[T1]{fontenc}
\usepackage{textcomp}
\usepackage[varqu,varl]{zi4}
\usepackage[all]{nowidow}
\usepackage[auth-lg,affil-it]{authblk}
\usepackage[keeplastbox]{flushend}
\usepackage{fancyhdr}

\usepackage{authblk}
\usepackage{booktabs}
\usepackage{comment}
\usepackage{tabularx}
\usepackage{tabularray}
\UseTblrLibrary{booktabs}
\usepackage{stfloats}
\usepackage{placeins}
\usepackage{hyperref}
\usepackage{multirow}

\fancypagestyle{firstpage}{
  \fancyhf{}
  \renewcommand{\headrulewidth}{0pt}
  \fancyfoot[C]{\thepage}
}

\begin{document}


\title{Performance Characterization of SPEC CPU\textregistered2026\\on AMD EPYC\texttrademark\space 9755 Processor}




\author{Kunal Kashyap}
\author{Rajiv Ramanathan}
\author{Shayantika Bhattacharya}
\affil{Server Power and Performance Optimization\\Advanced Micro Devices, Inc. (AMD)\\\{kunal.kashyap, rajiv.ramanathan, shayantika.bhattacharya\}@amd.com}

\maketitle
\thispagestyle{firstpage}
\pagestyle{plain}

\thispagestyle{fancy}
\fancyhf{}
\fancyfoot[C]{\footnotesize © 2026 IEEE. Personal use of this material is permitted. Permission from IEEE must be obtained for all other uses, in any current or future media, including reprinting/republishing this material for advertising or promotional purposes, creating new collective works, for resale or redistribution to servers or lists, or reuse of any copyrighted component of this work in other works.}
\renewcommand{\headrulewidth}{0pt}

\begin{abstract}
SPEC CPU 2026 is the first major update to the industry-standard CPU benchmark suite since 2017. This paper presents the first microarchitecture based performance characterization of the new suite, conducted on AMD EPYC "Zen 5", also the first SPEC CPU characterization study on this microarchitecture. Using a multi-lens methodology spanning pipeline efficiency, control flow behavior, cache hierarchy pressure, and instruction mix, we analyze both SPECrate and SPECspeed suites. We introduce scale analysis, comparing single-copy to full-system behavior to expose system-level bottlenecks invisible to conventional characterization. Our analysis reveals substantial behavioral diversity across the suite, and the multi-lens analysis identifies three distinct behavioral clusters: frontend control-flow-dominated workloads that stress branch predictor throughput rather than accuracy, high-efficiency compute workloads that suffer SMT contention at scale, and memory bandwidth-bound workloads with poor L3 filtering even at single-copy. Scale-dependent effects, including SMT dispatch contention causing throughput reduction and L3 capacity interference, emerge only at full system utilization. This work establishes an empirical foundation for architectural research and workload-driven design decisions targeting next-generation datacenter processors.
\end{abstract}

\section{Introduction}

Every generation of processor microarchitecture is shaped, in part, by the workloads used to evaluate it. Workload characterization studies reveal how applications stress hardware resources, guiding optimization decisions from compiler instruction scheduling to cache hierarchy design. Standard benchmark suites provide the reproducible reference points that make such studies comparable across systems and lasting across years. SPEC CPU\textregistered\space has served as the industry-standard benchmark suite for CPU performance evaluation for over three decades, with each iteration reflecting the evolving computational demands of its era.

Two characterization gaps motivate this work. First, SPEC CPU\textregistered2026~\cite{speccpu2026}, the first major update to the suite since 2017, lacks a systematic microarchitectural characterization on any platform. Second, prior SPEC CPU studies have focused predominantly on Intel platforms~\cite{hebbar2019,Kejariwal2008ComparativeAC,limaye2018,Singh_2019,navarro-torres2019}; no SPEC CPU suite has been characterized on AMD "Zen 5", or to our knowledge, on any "Zen" microarchitecture, despite AMD EPYC\texttrademark\space processors~\cite{amdepycprocs} now powering a substantial share of datacenter infrastructure~\cite{amdmarketshare}. This paper addresses both gaps.

Conventional workload characterization studies profile single benchmark instances. However, server deployments run hundreds of concurrent workloads, and behavior at scale can diverge substantially from single-instance profiles. We introduce a scale analysis methodology, comparing single-instance to full-system behavior. This approach exposes system-level bottlenecks invisible to conventional characterization: cache capacity interference patterns, SMT dispatch contention, and memory bandwidth saturation that emerge only when the full system is utilized. Understanding these scale-dependent effects is critical for server-class processor deployment.

This paper makes the following contributions:
\begin{itemize}
    \item \textbf{First microarchitectural characterization of SPEC CPU\textregistered2026} using a multi-lens methodology (pipeline efficiency, control flow behavior, cache hierarchy pressure, instruction mix) on AMD EPYC\texttrademark.
    \item \textbf{Scale analysis methodology} comparing single-instance to full-system behavior, exposing bottlenecks invisible to conventional workload characterization.
    \item \textbf{Workload taxonomy} identifying three behavioral clusters based on dominant bottleneck type, with distinct optimization implications.
    \item \textbf{Parallelism analysis} of SPECspeed\textregistered2026 validating the suite's expanded multi-threaded coverage and revealing fundamental scaling diversity across application domains.
\end{itemize}

The remainder of this paper is organized as follows: Section \ref{background} provides background on SPEC CPU 2026, ``Zen~5'' microarchitecture, and related work. Section \ref{methodology} describes our experimental methodology. Section \ref{characterization} presents the four-lens workload characterization. Section \ref{synthesis} synthesizes findings into a workload taxonomy and ``Zen~5''-specific observations. Section \ref{conclusion} concludes with a summary of contributions and directions for future work.

\section{Background and Related Work}
\label{background}

\subsection{SPEC CPU\textregistered2026 Overview}
SPEC CPU\textregistered2026 is the latest iteration of the industry-standard CPU benchmark suite \cite{specisca}, released in May 2026. The suite comprises four benchmark suites, containing 52 applications in total, targeting different workload characteristics: SPECrate\textregistered2026 Integer and SPECrate\textregistered2026 Floating Point for throughput-oriented multi-copy workloads, and SPECspeed\textregistered2026 Integer and SPECspeed\textregistered2026 Floating Point for latency-oriented parallel workloads. Each benchmark includes three workload sizes: test, train, and reference (ref), with test being the smallest and ref the largest. The reference workloads are used for official timing and scoring.

Compared to its predecessor, SPEC CPU\textregistered2017, the 2026 suite elicits a broader set of microarchitecture responses and introduces several new domains including cryptography (750.sealcrypto), neuroscience (767.nest) and graph analytics (854.graph500). The SPECspeed\textregistered2026 Integer suite addresses a critical gap from CPU 2017 with the inclusion of multiple (9 out of 13) multi-threaded integer benchmarks, compared to only one in CPU 2017 \cite{specisca}.

\subsection{5th Gen AMD EPYC\texttrademark \space Processor Architecture}
\label{zen5}
The 5th Gen AMD EPYC\texttrademark \space processors (formerly codenamed "Turin") employ a hybrid multi-die architecture that decouples CPU cores from I/O functions~\cite{amd2025epycwhitepaper}. This design combines multiple Core Complex Dies (CCDs) with a central I/O die, enabling AMD to fabricate CPU cores on leading-edge process nodes while using mature processes for analog I/O circuitry.

``Zen~5'' is a ground‑up redesign of AMD’s frontend and execution engine on TSMC 4\,nm~\cite{singhZen52025,amd2024hotchips}. Each CCD integrates 8 cores with a shared 32\,MB L3 cache; our experimental setup uses the standard ``Zen~5'' cores (not the density‑optimized ``Zen~5c'').

\begin{figure}[bp]
\centerline{\includegraphics[width=\columnwidth]{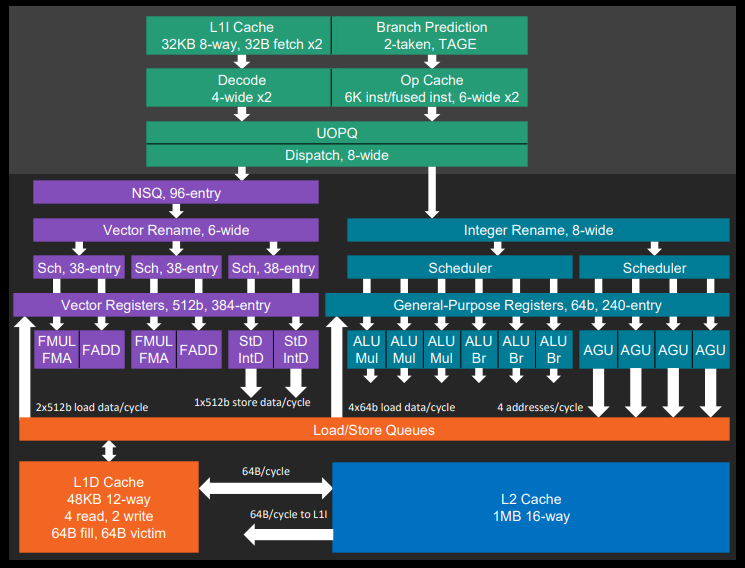}}
\caption{"Zen 5" Microarchitecture.}
\label{zen5_uarch}
\end{figure}

Figure 1 summarizes the microarchitecture. The frontend widens substantially, combining dual‑pipe instruction fetch from a 32\,KB L1 I‑cache with a large 6K‑entry Op Cache capable of supplying up to 12 fused micro‑ops per cycle. When decoding is required, legacy decoders scale to 8 instructions per cycle. Branch prediction is reinforced via a much larger L1 BTB and a TAGE‑based predictor, improving accuracy for deep speculation.

``Zen~5'' adopts a unified scheduling model with an 8‑wide dispatch/retire path. Integer execution is backed by a large physical register file and multiple ALUs, multipliers, branch units, and AGUs, enabling high instruction‑level parallelism. Floating‑point and SIMD execution move to native 512‑bit AVX‑512 datapaths, eliminating double‑pumping and allowing single‑cycle 512‑bit loads with multiple FP pipelines in flight~\cite{amd2025epycwhitepaper}.

The memory hierarchy is scaled accordingly. The L1 data cache grows to 48\,KB while sustaining two 512‑bit loads and one 512‑bit store per cycle, L1–L2 bandwidth doubles to 64\,B/cycle, and each core retains a private 1\,MB, 16‑way L2 cache. At the socket level, a 6\,nm I/O die integrates 12 DDR5‑6400 controllers delivering up to 614\,GB/s per socket, with Infinity Fabric links connecting CCDs to memory and I/O. Together, these changes align frontend throughput, execution width, and memory bandwidth, enabling ``Zen~5'' to sustain higher compute and data rates across diverse workloads.

\vspace{-4pt}
\subsection{Related Work}
Benchmark characterization has been used for decades to guide architectural design \cite{Conte1991BenchmarkC}. These studies have accompanied each major SPEC CPU release, providing essential insights for architecture research and compiler optimization. For SPEC CPU\textregistered\space2006, studies were conducted during development to discover hot function routines~\cite{weicker2007}, to highlight challenges and insights derived from event-based analysis~\cite{henning2007}, to compare suite versions~\cite{Kejariwal2008ComparativeAC}, and to perform statistical analysis for benchmark subsetting~\cite{phansalkar2007}. Hassan, Park, and Black-Schaffer~\cite{hassan2021} later provided a reusable, microarchitecture-independent characterization of memory system behavior across both CPU 2006 and CPU 2017.

Following the release of CPU 2017, Limaye and Adegbija~\cite{limaye2018} published a comprehensive workload characterization covering instruction mix, branch behavior, and cache characteristics, with explicit comparison to CPU2006. Panda, Song, Dean, and John~\cite{panda2018} asked whether the new suite broadened the performance horizon, finding that CPU 2017 workloads exhibit significantly higher dynamic instruction counts and stress memory hierarchies more heavily than their predecessors. Academic studies have provided memory-centric characterization of CPU 2017, detailing memory footprints and bandwidth patterns~\cite{Singh_2019}, and evaluating memory hierarchy response on Intel Skylake~\cite{navarro-torres2019}. Hebbar and Milenkovi\'{c}~\cite{hebbar2019} combined top-down microarchitectural analysis with energy metrics on the Core i7-8700K. These studies collectively establish that CPU 2017 shifted toward more memory-intensive and longer-running workloads, a trend we expect CPU 2026 to continue. 

While "Zen 5's" microarchitecture has been documented through industry publications (Section \ref{zen5}), systematic SPEC CPU characterization on this architecture remains absent.

\section{Methodology}
\label{methodology}

\subsection{Benchmark Configuration}
All benchmarks are compiled using \texttt{GCC 15.2.0} with optimization flags \texttt{-O3 -march=znver5 -flto}. The explicit \texttt{-march=znver5} flag ensures consistent targeting of the full "Zen 5" ISA, including AVX-512 and VNNI extensions. Link-time optimization (LTO) is enabled to allow cross-module inlining and optimization, consistent with performance-oriented compilation practices. Compiler and optimization choices affect benchmark characteristics; our configuration represents a realistic high-performance setting without aggressive transformations (such as profile-guided optimization) that could obscure inherent workload behavior.

All experiments use the reference (ref) workloads, which are the largest input sets and are used for official SPEC CPU scoring. We collect data for both rate and speed suites to provide a comprehensive characterization.
\begin{itemize}
    \item \textbf{Rate:} We run 1-copy and 512-copy configurations. Single copy runs isolate per-workload behavior without inter-copy interference or memory bandwidth bottlenecks, enabling clean microarchitectural profiling. The 512-copy configuration (matching the hardware's logical thread count) represents maximum throughput operation and reveals scalability characteristics and resource contention patterns.
    \item \textbf{Speed:} We run 1-thread and 512-thread configurations. Single thread runs establish baseline sequential performance, while 512-thread runs expose the parallelization characteristics of each benchmark. Notably, not all speed benchmarks scale to the full thread count as some are inherently serial or have limited parallel regions, making the utilization differential a key characterization metric.
\end{itemize}

\subsection{Experimental Setup}
All data is collected on a system built around an AMD EPYC\texttrademark \space9005 Series processor featuring the "Zen 5" core. Table \ref{tab:expsetup} summarizes the full hardware and software operating environment.

\begin{table}[b]                                                                                                   \centering
\caption{Experimental Setup}                                                                                                           
\label{tab:expsetup}
\begin{tabular}{@{}ll@{}}                                                                                                                               
\toprule                                                                                                                                                
\textbf{Component} & \textbf{Configuration} \\
\midrule
Processor          & AMD EPYC\texttrademark \space9755 \\
Frequency          & 2.7 GHz (Max. Boost to 4.1 GHz$^\dagger$) \\
Sockets            & 2 \\
\midrule
L1 Cache           & 32 KiB I + 48 KiB D per core \\
L2 Cache           & 1 MiB per core \\
L3 Cache           & 512 MiB \\
\midrule
Memory             & 2.3 TB DDR5-6400 \\
\midrule
Operating System   & Ubuntu 24.04 LTS \\
Kernel             & 6.8.0-44-generic \\
Compiler           & GCC 15.2.0 \texttt{-O3 -march=znver5 -flto} \\
\bottomrule
\end{tabular}
\vspace{2pt}
\parbox{\linewidth}{\scriptsize $^\dagger$EPYC-18: Max boost for AMD EPYC\texttrademark\space processors is the maximum frequency achievable by any single core on the processor under normal operating conditions for server systems.}
\end{table}

\subsection{Performance Measurement Framework}
\label{tools}

\textbf{Profiling Tool.} We use AMD uProf for all performance counter collection, including the metrics for Top-Down Microarchitectural analysis. uProf is a publically available perfomance analysis tool~\cite{uprof} that provides access to the full set of AMD-specific PMU events, including pipeline utilization, instruction mix, detailed cache hierarchy and memory bandwidth metrics. The \texttt{AMDuProfPcm} executable supports system-wide time series sampling, which we configure to collect approximately 100 metrics per sample at 2-second intervals. 

For rate benchmarks, we filter samples to those where CPU utilization exceeds 95\%, ensuring only steady-state execution phases are included in the analysis. For speed benchmarks with limited parallelism, we include all samples to capture true utilization characteristics.

\textbf{Top-Down Microarchitecture Analysis (TMA).} We adopt the Top-Down Microarchitecture Analysis methodology\cite{yasintma} to classify pipeline slot utilization. TMA partitions execution into four top-level categories:
\begin{itemize}
    \item \textbf{Frontend Bound:} Cycles where the backend is ready but the frontend cannot deliver micro-ops, indicating instruction fetch or decode bottlenecks.
    \item \textbf{Backend Bound:} Cycles where the frontend has delivered micro-ops but the backend cannot accept them, typically due to execution unit or memory subsystem stalls.
    \item \textbf{Pipeline Stalls:} Pipeline slots wasted due to branch mispredictions or machine clears.
    \item \textbf{Retiring:} Pipeline slots that result in useful work (committed instructions).
\end{itemize}

AMD uProf provides TMA-compatible metrics through its `pipeline\_util` preset, which we use to compute these categories. In addition to the four top-level categories, it also provides a fifth category, \textit{SMT-contention}, which signifies the percentage of unused dispatch slots as the other thread was selected. This framework enables systematic comparison across workloads and identifies the primary performance limiters for each benchmark.
\textbf{Counter Multiplexing.} Collecting the full set of almost 100 metrics requires multiplexing across multiple PMU counter groups, as the hardware provides a limited number of concurrent counters. \texttt{AMDuProfPcm} handles this transparently through time-sliced sampling. For long-running benchmarks (typical SPEC CPU 2026 workloads can run anywhere from 3 to 60 minutes), statistical averaging across many samples yields stable metric values with low variance.

\section{Microarchitectural Characterization of SPEC CPU\textregistered2026}
\label{characterization}

We characterize SPEC CPU\textregistered2026 workloads through four complementary lenses, each answering a distinct question about benchmark behavior:

\begin{itemize}
    \item \textbf{Top-Down Microarchitectural Analysis:} Where are the pipeline slots spending time, and how efficiently is the core being utilized?
    \item \textbf{Control Flow Characteristics:} How does control flow complexity and branch behavior vary across workloads?
    \item \textbf{Cache and Memory Subsystem Behavior:} Where in the memory hierarchy do workloads apply pressure?
    \item \textbf{Instruction Mix and SIMD Utilization:} What does instruction mix reveal about computational character and vectorization behavior?
\end{itemize}

Each subsection presents data from AMD uProf profiling, highlighting clusters of similar workloads and outliers that merit attention. Cross-references connect observations across lenses, building toward the synthetic analysis in Section \ref{synthesis}.

\subsection{Top-Down Microarchitectural Analysis}
\label{tmam}

Using the TMA framework introduced in Section \ref{tools}, we characterize where pipeline slots are consumed across CPU 2026 benchmarks.

\begin{figure*}[b]
  \centering
  \begin{minipage}[t]{0.48\textwidth}
    \centering
    \includegraphics[width=\linewidth]{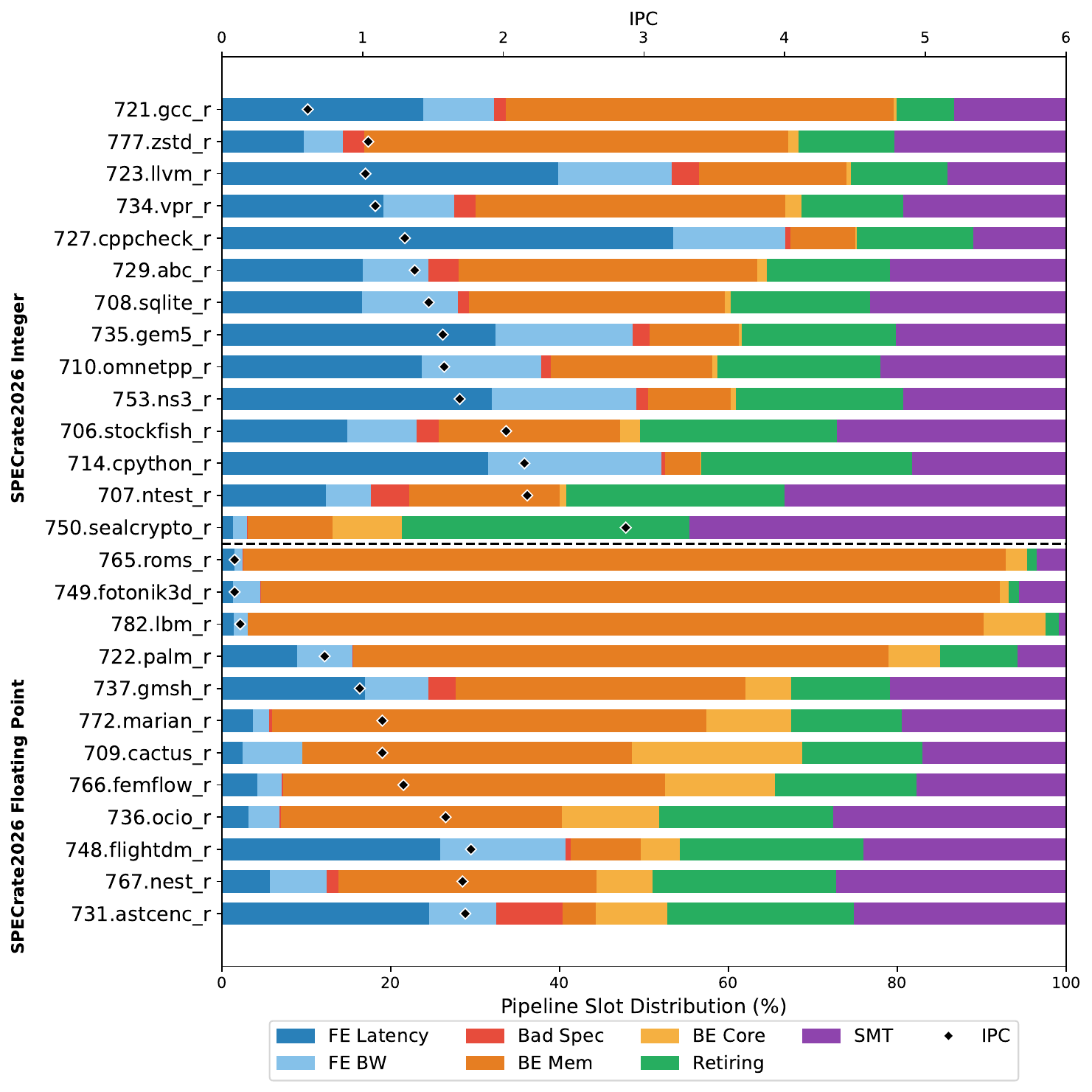}
    \caption{TMA Breakdown of 512-copy Rate Benchmarks.}
    \label{tma1}
  \end{minipage}\hfill
  \begin{minipage}[t]{0.48\textwidth}
    \centering
    \includegraphics[width=\linewidth]{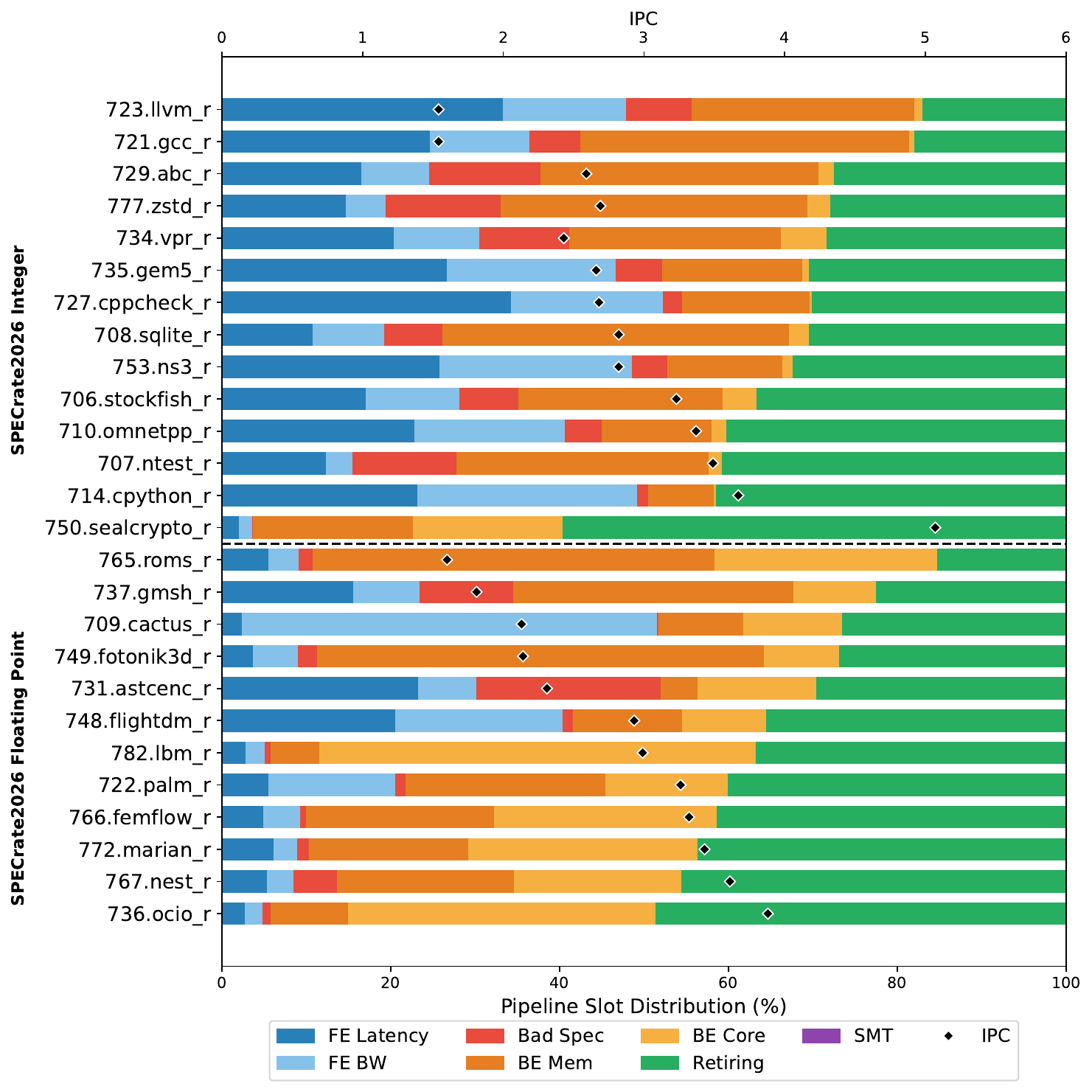}
    \caption{TMA Breakdown of 1-copy Rate Benchmarks.}
    \label{tma2}
  \end{minipage}
\end{figure*}

In Fig.~\ref{tma1}-\ref{tma4} each stacked bar represents the distribution of pipeline slots across seven categories. Frontend (FE) stalls are decomposed into FE Latency (instruction cache misses, iTLB misses, branch resteer delays) and FE Bandwidth (decode throughput limitations). Bad Speculation represents slots wasted on mispredicted paths. Backend (BE) stalls are decomposed into BE Memory (cache/memory latency and bandwidth constraints) and BE Core (execution unit contention, serializing operations). Retiring represents useful work completing. Finally, SMT Contention captures dispatch slot competition between SMT threads sharing a physical core. Black markers on the top axis indicate Instructions Per Cycle (IPC) for each benchmark, allowing direct comparison between the TMA bottleneck distribution and resulting throughput.

\subsubsection{SPECrate\textregistered2026 at 512-Copy (Fig.~\ref{tma1})} Rate benchmarks are throughput metrics, and 512-copy execution, utilizing all hardware threads, represents realistic deployment. At this scale, system-level bottlenecks dominate the TMA breakdown.

The stencil-computation workloads 765.roms\_r and 749.fotonik3d\_r, along with the lattice-based 782.lbm\_r show severe memory subsystem saturation, with Retiring collapsed to ~1\%. These workloads, along with 722.palm\_r and 772.marian\_r, represent the most memory-constrained benchmarks in the suite.

We observe that SMT dispatch contention averages 19.3\% across the suite at 512-copy. The highest SMT contention occurs in 750.sealcrypto\_r, 707.ntest\_r, and 736.ocio\_r. These workloads have high dispatch efficiency, making them particularly sensitive to slot competition when two threads share a physical core.

The highest Retiring at 512-copy are 750.sealcrypto\_r, 767.nest\_r, and 736.ocio\_r, all showing substantial SMT contention that limits further efficiency gains.

IPC at 512-copy ranges from 0.42 (765.roms\_r) to 2.71 (750.sealcrypto\_r), reflecting severe throughput degradation under SMT and memory pressure. The correlation between Retiring and IPC is evident: workloads with higher Retiring percentages generally achieve higher IPC, as both metrics reflect useful work completion. However, the relationship is not linear; 750.sealcrypto\_r achieves 2.71 IPC despite only 34.0\% Retiring because its high dispatch efficiency means each retiring instruction represents more work. Memory-bound workloads like 765.roms\_r, 749.fotonik3d\_r, and 782.lbm\_r show IPC below 0.5, consistent with their near-complete backend stall.

Comparing across suites, FP benchmarks are predominantly backend memory-bound at 512-copy, with the five most memory-constrained workloads all belonging to FP Rate. In contrast, integer benchmarks exhibit more balanced TMA profiles, with higher representation in the frontend bound and SMT-contention categories.

\subsubsection{SPECrate\textregistered2026 at 1-Copy (Fig.~\ref{tma2})} Single-copy execution isolates workload-specific microarchitectural behavior without system-level contention, revealing inherent bottleneck characteristics.

The highest Retiring at 1-copy are 750.sealcrypto\_r, 736.ocio\_r, 767.nest\_r, 772.marian\_r, and 766.femflow\_r. These workloads effectively utilize the "Zen 5" core's wide execution resources.

Frontend-dominated workloads indicate instruction supply bottlenecks. The highest Frontend Bound are 727.cppcheck\_r, 709.cactus\_r, 753.ns3\_r, 714.cpython\_r, and 723.llvm\_r. These workloads are characterized by complex control flow patterns that stress the branch prediction and instruction fetch subsystems.

Backend Memory-dominated workloads at 1-copy show inherent memory intensity before system saturation. The highest BE Memory are 749.fotonik3d\_r, 765.roms\_r, 708.sqlite\_r, 721.gcc\_r, and 777.zstd\_r.

IPC at 1-copy spans from 1.54 (721.gcc\_r, 723.llvm\_r) to 5.07 (750.sealcrypto\_r), a 3.3x range reflecting the diversity of computational patterns. The TMA breakdown explains this variation: 750.sealcrypto\_r achieves 5.07 IPC with 59.7\% Retiring, while 721.gcc\_r manages only 1.54 IPC with 36.3\% Frontend Bound. The highest IPC workloads correlate with high Retiring, though exceptions reveal that Retiring alone does not guarantee high IPC; 714.cpython\_r achieves 3.67 IPC despite 48.4\% Frontend Bound because its frontend stalls are latency-tolerant.

Memory-bound workloads show lower IPC despite moderate Retiring: 765.roms\_r achieves only 1.60 IPC because its backend memory stalls (45.3\%) create long-latency pipeline bubbles. The integer suite averages 2.82 IPC versus 2.79 for FP, a surprising parity explained by integer's higher frontend pressure being offset by FP's higher memory pressure.

This INT/FP divergence persists at 1-copy: FP benchmarks cluster toward backend memory constraints (749.fotonik3d\_r, 765.roms\_r lead BE Memory), while integer benchmarks dominate the frontend bound category (727.cppcheck\_r, 753.ns3\_r, 714.cpython\_r, 723.llvm\_r). The pattern reflects fundamental differences in computational structure: integer workloads feature complex control flow and pointer-heavy data structures, while FP workloads operate on large numerical arrays with regular access patterns.

Comparing 1-copy to 512-copy reveals systematic scale-dependent shifts in bottleneck profiles. Backend Memory amplifies dramatically for memory-intensive workloads: 765.roms\_r increases from 45.3\% to 90.3\%, and 749.fotonik3d\_r from 54.0\% to 87.5\%. Correspondingly, Retiring collapses as memory bandwidth saturates. SMT dispatch contention, absent at 1-copy, emerges as a significant factor at 512-copy (averaging 19.3\%), particularly impacting workloads with high dispatch efficiency. Frontend Bound percentages generally compress as backend constraints dominate at scale.

\textbf{Comparison with SPEC CPU\textregistered2017:} To contextualize SPEC CPU\textregistered2026's behavioral scope, we compare 1-copy IPC distributions between CPU 2026 and CPU 2017 on the same ``Zen~5'' system, with both suites built using the identical toolchain and flags of Table~\ref{tab:expsetup}. All reported averages are computed over the full set of benchmarks in each respective suite.

SPEC CPU\textregistered2026 Integer suite shows a higher average and median IPC than SPEC CPU\textregistered2017 Integer (2.90 vs. 2.53 and 2.76 vs. 2.21 respectively, Table~\ref{tab:ipc_compare}), reflecting the inclusion of workloads that more effectively utilize ``Zen~5's'' wide execution resources. CPU 2026 FP exhibits a notably compressed IPC distribution, i.e., a higher floor (1.60 vs. 1.03) with a narrower range (2.28 vs. 3.93), indicating more homogeneous memory-bound behavior across the floating-point suite.

Beyond IPC, our TMA data also point to a qualitative shift within the frontend category. The overall frontend- versus backend-bound balance is comparable across the two suites, but several SPEC CPU\textregistered2026 benchmarks appear to carry a larger share of frontend stalls from fetch latency: instruction-cache and iTLB misses, rather than decode bandwidth, consistent with the larger, more complex code footprints of the updated suite. We report this as a preliminary, qualitative observation: a rigorous frontend decomposition was beyond the scope of this study, and we do not present the supporting breakdown here. We leave a detailed characterization of this frontend shift to future work.

\begin{table}[bp]
\centering
\caption{1-copy IPC Comparison\\SPECrate\textregistered2026 vs.\ SPECrate\textregistered2017}
\label{tab:ipc_compare}
\setlength{\tabcolsep}{4pt}
\begin{tabular}{@{}lrrrrr@{}}
\toprule
\textbf{Suite} & \textbf{Max} & \textbf{Min} & \textbf{Avg} & \textbf{Median} & \textbf{Range} \\
\midrule
SPECrate\textregistered2026 Integer & 5.07 & 1.54 & 2.90 & 2.76 & 3.53 \\
SPECrate\textregistered2017 Integer & 4.09 & 1.38 & 2.53 & 2.21 & 2.71 \\
\midrule
SPECrate\textregistered2026 FP  & 3.88 & 1.60 & 2.78 & 2.96 & 2.28 \\
SPECrate\textregistered2017 FP  & 4.96 & 1.03 & 2.59 & 2.60 & 3.93 \\
\bottomrule
\end{tabular}
\end{table}

\begin{figure*}[t]
  \centering
  \begin{minipage}[t]{0.48\textwidth}
    \centering
    \includegraphics[width=\linewidth]{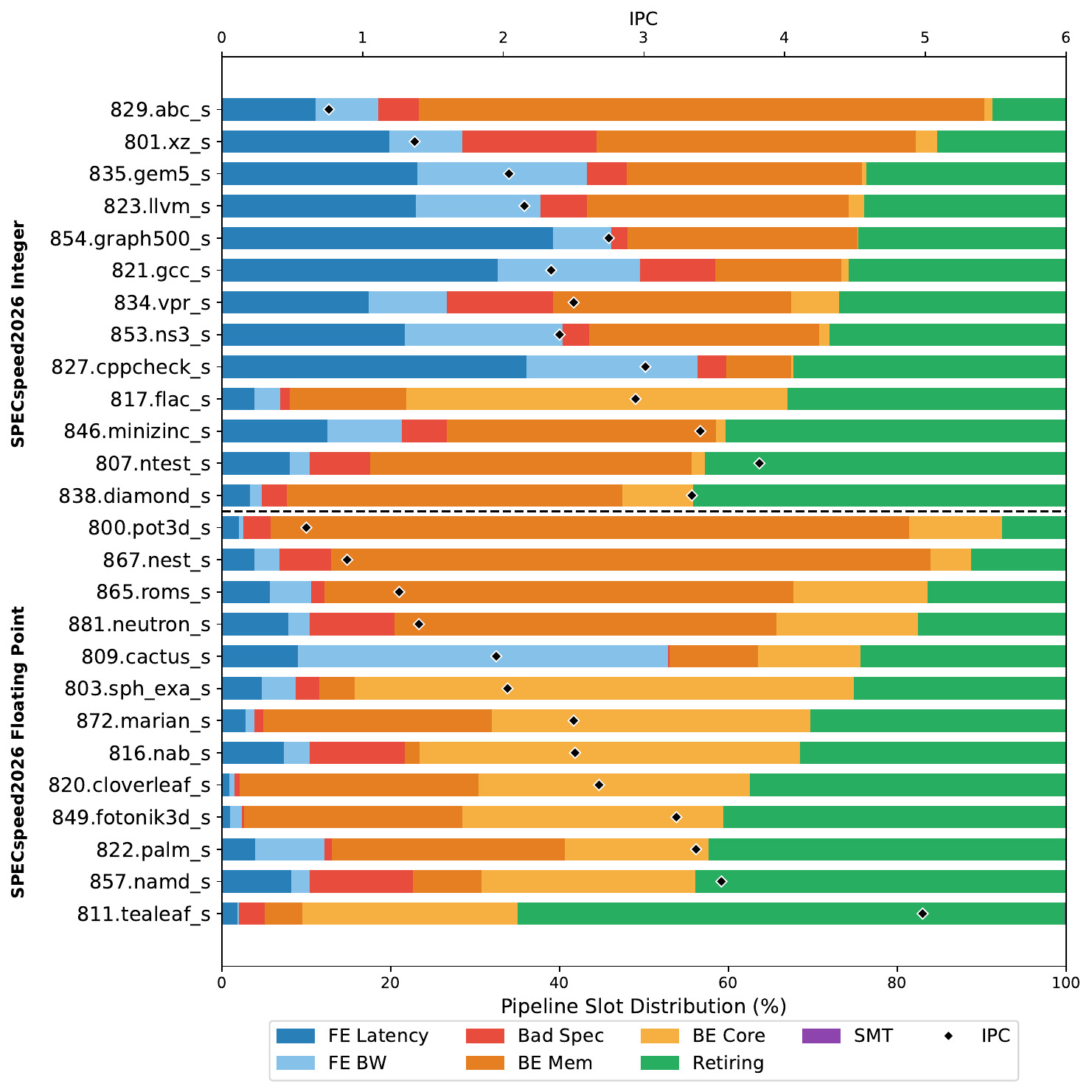}
    \caption{TMA Breakdown of 1-thread Speed Benchmarks.}
    \label{tma3}
  \end{minipage}\hfill
  \begin{minipage}[t]{0.48\textwidth}
    \centering
    \includegraphics[width=\linewidth]{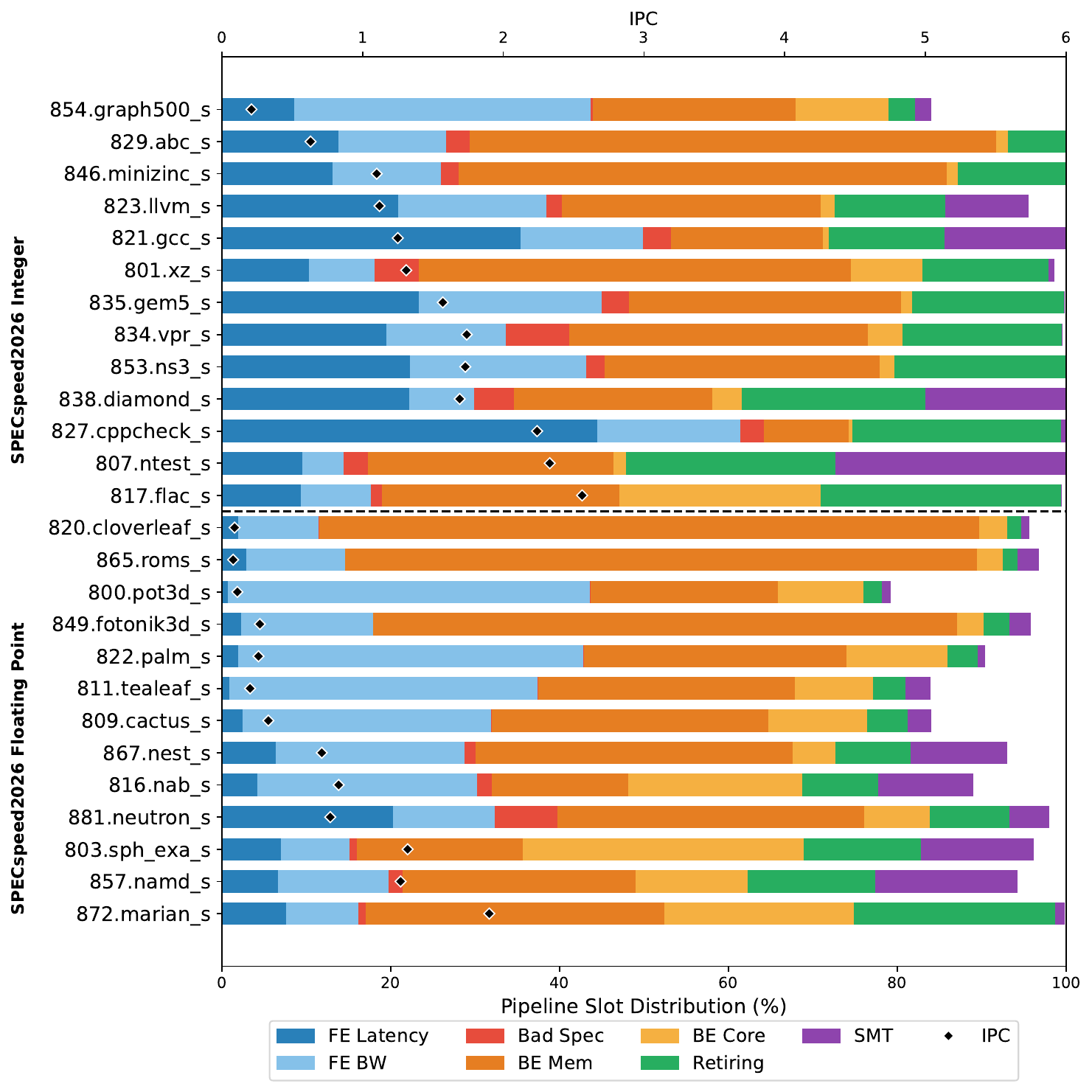}
    \caption{TMA Breakdown of 512-thread Speed Benchmarks.}
    \label{tma4}
  \end{minipage}
\end{figure*}

\subsubsection{SPECspeed\textregistered2026 at Single Thread (Fig.~\ref{tma3})} The speed benchmark suite exhibits different bottleneck distributions than rate, reflecting different input sizes and workload phases. TMA metrics for 512-thread speed benchmarks do not consistently sum to 100\% (Fig.~\ref{tma4}). While 512-copy rate benchmarks and 1-thread speed benchmarks yield complete accounting (99.9\% and 100.0\% respectively), 512-thread speed benchmarks average only 95.3\%, concentrated in HPC codes with barrier-heavy communication (800.pot3d\_s, 811.tealeaf\_s, 854.graph500\_s, 809.cactus\_s, 816.nab\_s). This reflects blocking synchronization (MPI barriers, OpenMP implicit barriers) which falls outside TMA's accounting framework~\cite{kleen2017pmutools,yasintma}. Rate benchmarks, running independent copies with no inter-process synchronization, exhibit no gap.

For speed benchmark characterization, we therefore present 1-thread TMA profiles and use utilization/speedup metrics for scaling analysis.

The most Backend Memory-dominated speed workloads are 800.pot3d\_s, 867.nest\_s, and 829.abc\_s. These represent electromagnetic propagation, neural simulation at scale, and logic synthesis respectively.

Frontend-dominated speed workloads include 827.cppcheck\_s, 809.cactus\_s, and 854.graph500\_s. The graph500\_s benchmark's frontend pressure stems from pointer-chasing graph traversal patterns.

The highest Retiring speed workloads are 811.tealeaf\_s, 838.diamond\_s, and 857.namd\_s, representing heat conduction simulation, sequence alignment, and molecular dynamics respectively.

IPC for speed benchmarks ranges from 1.26 (823.llvm\_s) to 4.59 (811.tealeaf\_s). The speed suite shows stronger IPC-Retiring correlation than rate benchmarks: 811.tealeaf\_s achieves both the highest Retiring (64.7\%) and highest IPC (4.59), while frontend-bound 827.cppcheck\_s shows 2.39 IPC with 56.2\% Frontend Bound. The memory-intensive 800.pot3d\_s achieves only 1.46 IPC due its 75.7\% Backend Memory, consistent with the rate benchmark observation that memory-bound workloads suffer IPC degradation regardless of scale.

\subsubsection{SPECspeed\textregistered2026 Parallelism (Table~\ref{tab:speed_parallel})} Unlike rate benchmarks where all copies execute identical workloads, speed benchmarks have varying inherent parallelism. These parallel benchmarks use one of the following techniques: OpenMP 3, C++'s \texttt{std::thread}, Fortran's \texttt{DO CONCURRENT}, or task-based process spawning \cite{specisca}. 

\begin{table}[tbp]
\centering
\caption{SPECspeed\textregistered2026 Benchmarks' Parallelism}
\label{tab:speed_parallel}
\begin{tabular}{@{}c lrr@{}}
\toprule
& \textbf{Benchmark} & \textbf{Util. 512t (\%)} & \textbf{Speedup} \\
\midrule
\multirow{12}{*}{\rotatebox[origin=c]{90}{\textbf{Integer}}}
& 854.graph500\_s & 91.8 & 21.5$\times$ \\
& 821.gcc\_s & 87.4 & 177.7$\times$ \\
& 807.ntest\_s & 81.9 & 189.1$\times$ \\
& 823.llvm\_s & 73.7 & 166.5$\times$ \\
& 838.diamond\_s & 47.1 & 116.9$\times$ \\
& 827.cppcheck\_s & 26.0 & 55.9$\times$ \\
& 801.xz\_s & 6.6 & 17.9$\times$ \\
& 817.flac\_s & 1.4 & 6.8$\times$ \\
& 846.minizinc\_s & 0.8 & 1.0$\times$ \\
& 829.abc\_s & 0.3 & 0.8$\times$ \\
& 834.vpr\_s & 0.3 & 0.9$\times$ \\
& 835.gem5\_s & 0.3 & 1.0$\times$ \\
& 853.ns3\_s & 0.3 & 1.0$\times$ \\
\midrule
\multirow{13}{*}{\rotatebox[origin=c]{90}{\textbf{Floating Point}}}
& 811.tealeaf\_s & 99.8 & 14.6$\times$ \\
& 820.cloverleaf\_s & 99.8 & 14.2$\times$ \\
& 865.roms\_s & 93.2 & 29.8$\times$ \\
& 816.nab\_s & 85.5 & 86.9$\times$ \\
& 867.nest\_s & 81.9 & 164.2$\times$ \\
& 849.fotonik3d\_s & 81.2 & 20.0$\times$ \\
& 822.palm\_s & 75.1 & 3.0$\times$ \\
& 857.namd\_s & 60.8 & 76.4$\times$ \\
& 803.sph\_exa\_s & 56.1 & 125.0$\times$ \\
& 881.neutron\_s & 40.9 & 43.3$\times$ \\
& 809.cactus\_s & 28.5 & 19.6$\times$ \\
& 800.pot3d\_s & 17.6 & 1.0$\times$ \\
& 872.marian\_s & 3.5 & 11.6$\times$ \\
\bottomrule
\end{tabular}
\end{table}

High parallelism benchmarks (>80\% utilization) include 820.cloverleaf and 811.tealeaf\_s, 865.roms\_s, 854.graph500\_s, 821.gcc\_s, 816.nab\_s, and 867.nest\_s/807.ntest\_s. Notably, 821.gcc\_s and 823.llvm\_s represent the first task-parallel benchmarks in SPEC CPU history, based on the two most widely-used open-source compilers, each invokes thousands of independent compilation commands to build multiple source files, enabling their high utilization through embarrassingly parallel workloads.

Limited parallelism benchmarks (<5\% utilization) include 872.marian\_s, 817.flac\_s, and several essentially serial workloads: 829.abc\_s, 853.ns3\_s, 835.gem5\_s, 834.vpr\_s and 846.minizinc\_s (<1\% each).

While the SPECspeed\textregistered2026 Integer suite now includes many more multi-threaded benchmarks compared to  CPU 2017, several benchmarks still remain largely serial, reflecting application domains where parallelization is algorithmically challenging or lack of representative workloads which would meet SPEC CPU's requirements.

\subsection{Control Flow Characteristics}
\label{controlflow}

Branch density and prediction accuracy shape instruction-level parallelism by determining how effectively the processor can speculate past control flow boundaries. Table \ref{tab:branch_compact} presents control flow metrics for all SPEC CPU\textregistered2026 benchmarks.

\begin{table}[tbp]
\centering
\caption{Branch Behavior Metrics\\(1-copy SPECrate\textregistered\space vs. 1-thread SPECspeed\textregistered)}
\label{tab:branch_compact}
\setlength{\tabcolsep}{3.0pt}
\renewcommand{\arraystretch}{1.03}
\begin{tabular}{@{}c l rr rr@{}}
\toprule
& \textbf{Benchmark} &
\multicolumn{2}{c}{\textbf{BrKI}} &
\multicolumn{2}{c}{\textbf{Mispred (\%)}} \\
\cmidrule(lr){3-4}\cmidrule(lr){5-6}
& & \textit{Rate} & \textit{Speed}
& \textit{Rate} & \textit{Speed} \\
\midrule

\multirow{14}{*}{\rotatebox[origin=c]{90}{\textbf{Integer}}}
& 706.stockfish\_r   & 103.8 & --    & 2.7 & --    \\
& 707.ntest\_r, 807.ntest\_s       & 81.2  & 49.1  & 2.7 & 2.6  \\
& 708.sqlite\_r      & 206.8 & --    & 0.9 & --    \\
& 710.omnetpp\_r     & 198.4 & --    & 0.5 & --    \\
& 714.cpython\_r     & 208.3 & --    & 0.1 & --    \\
& 721.gcc\_r, 823.gcc\_s         & 222.8 & 223.8 & 1.6 & 1.9  \\
& 723.llvm\_r, 823.llvm\_s       & 219.9 & 231.6 & 2.5 & 1.0  \\
& 727.cppcheck\_r, 827.cppcheck\_s    & 285.5 & 276.5 & 0.3 & 0.4  \\
& 729.abc\_r, 829.abc\_s         & 157.1 & 164.1 & 2.7 & 2.6  \\
& 734.vpr\_r, 834.vpr\_s         & 192.7 & 186.5 & 1.7 & 1.9  \\
& 735.gem5\_r, 835.gem5\_s        & 212.3 & 167.5 & 0.9 & 1.2  \\
& 750.sealcrypto\_r  & 36.3  & --    & 0.4 & --    \\
& 753.ns3\_r, 853.ns3\_s         & 221.1 & 221.6 & 0.5 & 0.4  \\
& 777.zstd\_r        & 116.9 & --    & 3.2 & --    \\

\midrule
\multirow{12}{*}{\rotatebox[origin=c]{90}{\textbf{Floating Point}}}
& 709.cactus\_r, 809.cactus\_s      & 9.1   & 8.0   & 0.00 & 0.00  \\
& 722.palm\_r, 822.palm\_s        & 40.3  & 51.7  & 0.4 & 0.3  \\
& 731.astcenc\_r     & 96.5  & --    & 6.3 & --    \\
& 736.ocio\_r        & 89.9  & --    & 0.1 & --    \\
& 737.gmsh\_r        & 175.3 & --    & 2.8 & --    \\
& 748.flightdm\_r   & 204.7 & --    & 0.1 & --    \\
& 749.fotonik3d\_r, 849.fotonik3d\_s  & 75.9  & 45.5  & 0.0 & 0.1  \\
& 765.roms\_r, 865.roms\_s        & 105.8 & 105.2 & 0.2 & 0.3  \\
& 766.femflow\_r    & 47.0  & --    & 0.3 & --    \\
& 767.nest\_r, 867.nest\_s        & 136.7 & 135.7 & 0.5 & 1.4  \\
& 772.marian\_r, 872.marian\_s      & 97.3  & 76.8  & 0.3 & 0.1  \\
& 782.lbm\_r         & 9.0   & --    & 0.2 & --    \\

\midrule
\multirow{5}{*}{\rotatebox[origin=c]{90}{\textbf{Integer}}}
& 801.xz\_s          & --    & 154.5 & --   & 5.7  \\
& 817.flac\_s        & --    & 104.2 & --   & 0.4  \\
& 838.diamond\_s     & --    & 54.5  & --   & 1.6  \\
& 846.minizinc\_s    & --    & 147.6 & --   & 0.8  \\
& 854.graph500\_s    & --    & 301.1 & --   & 1.0  \\
\midrule
\multirow{7}{*}{\rotatebox[origin=c]{90}{\textbf{Floating Point}}}
& 800.pot3d\_s       & --    & 73.3  & --   & 0.1  \\
& 803.sph\_exa\_s    & --    & 128.2 & --   & 0.6  \\
& 811.tealeaf\_s     & --    & 41.2  & --   & 0.00  \\
& 816.nab\_s         & --    & 133.5 & --   & 2.0  \\
& 820.cloverleaf\_s  & --    & 67.2  & --   & 0.0  \\
& 857.namd\_s        & --    & 35.5  & --   & 4.8  \\
& 881.neutron\_s     & --    & 116.3 & --   & 6.1  \\

\bottomrule
\end{tabular}

\vspace{2pt}
{\scriptsize \textit{Note:} BrKI denotes branches per thousand instructions. Empty cells indicate that the workload does not have a corresponding rate or speed variant in SPEC CPU 2026.}
\end{table}

The highest branch density workloads are 727.cppcheck\_r, 721.gcc\_r, 753.ns3\_r, 723.llvm\_r, and 735.gem5\_r. These compiler, static analysis, and simulation workloads exhibit complex control flow with one branch every 3-5 instructions.

The highest misprediction rates occur in 731.astcenc\_r, 777.zstd\_r, 737.gmsh\_r, 729.abc\_r, and 706.stockfish\_r. The 731.astcenc\_r (texture encoding) benchmark exhibits inherently unpredictable branch patterns in adaptive texture compression algorithms. 777.zstd\_r also exhibits control flow irregularity, which is typical in data compression.

727.cppcheck\_r's frontend bottleneck (Section \ref{tmam}) correlates with its exceptionally high branch density. This observation aligns with cppcheck’s role as a static analysis tool~\cite{cppcheck} whose execution is dominated by repeated character-level pattern matching across a large number of short token comparisons. With nearly one branch every 3-4 instructions, even with near-perfect prediction accuracy (0.32\% misprediction), the branch predictor throughput becomes the limiting factor, as each predicted branch consumes predictor bandwidth regardless of outcome. This represents a workload characteristic unlikely to improve through microarchitectural enhancements.

In contrast, 721.gcc\_r achieves comparable branch density (222.8 BrKI) but with higher misprediction (1.60\%), resulting in Backend Bound rather than Frontend Bound behavior. The difference lies in gcc's irregular data-dependent branching (pointer chasing, type dispatch) versus cppcheck's regular control flow patterns.

Comparing suites, integer benchmarks exhibit significantly higher branch density (mean 176 BrKI vs 91 BrKI for FP), consistent with their control-flow-intensive nature. 

Speed benchmarks exhibit similar branch characteristics to their rate counterparts. The highest branch density speed benchmarks are 854.graph500\_s, 827.cppcheck\_s, and 823.llvm\_s, all control-flow-intensive workloads. The highest misprediction rates occur in 881.neutron\_s, 801.xz\_s, and 857.namd\_s, reflecting Monte Carlo simulation, compression, and molecular dynamics with data-dependent branching.

\subsection{Cache and Memory Subsystem Behavior}
\label{cache&mem}

Cache hierarchy behavior determines whether workloads can exploit on-chip data locality or become constrained by off-chip memory bandwidth and latency. Table \ref{tab:cache_merged} presents cache hierarchy and memory bandwidth metrics for all SPEC CPU\textregistered2026 benchmarks in 1-copy and 512-copy modes respectively.

\begin{table*}[tbp]
\centering
\caption{Cache Hierarchy Metrics: Single-Copy/Thread vs. Full-Scale}
\label{tab:cache_merged}
\setlength{\tabcolsep}{2.2pt}
\renewcommand{\arraystretch}{1.03}
\begin{tabular}{@{}c l rr rr rr rr rr rr rr rr@{}}
\toprule
& & \multicolumn{8}{c}{\textbf{1-copy / 1-thread}} & \multicolumn{8}{c}{\textbf{512-copy / 512-thread}} \\
\cmidrule(lr){3-10}\cmidrule(lr){11-18}
& \textbf{Benchmark} &
\multicolumn{2}{c}{L1D MPKI} &
\multicolumn{2}{c}{L2 MPKI} &
\multicolumn{2}{c}{L3 MPKI} &
\multicolumn{2}{c}{Mem. BW} &
\multicolumn{2}{c}{L1D MPKI} &
\multicolumn{2}{c}{L2 MPKI} &
\multicolumn{2}{c}{L3 MPKI} &
\multicolumn{2}{c}{Mem. BW} \\
\cmidrule(lr){3-4}\cmidrule(lr){5-6}\cmidrule(lr){7-8}\cmidrule(lr){9-10}
\cmidrule(lr){11-12}\cmidrule(lr){13-14}\cmidrule(lr){15-16}\cmidrule(lr){17-18}
& & \textit{Rate} & \textit{Speed}
& \textit{Rate} & \textit{Speed}
& \textit{Rate} & \textit{Speed}
& \textit{Rate} & \textit{Speed}
& \textit{Rate} & \textit{Speed}
& \textit{Rate} & \textit{Speed}
& \textit{Rate} & \textit{Speed}
& \textit{Rate} & \textit{Speed} \\
\midrule

\multirow{14}{*}{\rotatebox[origin=c]{90}{\textbf{Integer}}}
& 706.stockfish\_r    & 20.6 & --   & 3.5  & --   & 0.3  & --   & 0.4  & --   & 27.6 & --   & 6.8  & --   & 1.7  & --   & 430  & --   \\
& 707.ntest\_r, 807.ntest\_s       & 3.0  & 7.4  & 0.4  & 0.7  & 0.0  & 0.0  & 0.0  & --   & 11.7 & 14.1 & 0.4  & 1.0  & 0.1  & 0.1  & 25   & 23   \\
& 708.sqlite\_r       & 8.9  & --   & 5.8  & --   & 1.3  & --   & 1.1  & --   & 9.2  & --   & 5.2  & --   & 2.1  & --   & 454  & --   \\
& 710.omnetpp\_r      & 12.2 & --   & 4.3  & --   & 0.1  & --   & 0.1  & --   & 24.7 & --   & 5.9  & --   & 1.0  & --   & 217  & --   \\
& 714.cpython\_r      & 0.9  & --   & 0.5  & --   & 0.2  & --   & 0.3  & --   & 7.2  & --   & 0.3  & --   & 0.2  & --   & 75   & --   \\
& 721.gcc\_r, 821.gcc\_s         & 22.1 & 13.3 & 17.5 & 5.0  & 3.9  & 0.3  & 1.9  & 0.2  & 29.0 & 15.8 & 19.6 & 6.6  & 6.2  & 1.9  & 682  & 334  \\
& 723.llvm\_r, 823.llvm\_s        & 17.6 & 12.5 & 18.1 & 8.8  & 1.8  & 2.1  & 0.8  & 1.4  & 16.3 & 8.9  & 7.8  & 5.5  & 2.3  & 2.5  & 375  & 272  \\
& 727.cppcheck\_r, 827.cppcheck\_s    & 16.0 & 14.8 & 8.4  & 8.7  & 4.8  & 2.9  & 3.5  & 2.4  & 14.7 & 8.8  & 9.3  & 4.4  & 4.4  & 3.4  & 808  & 208  \\
& 729.abc\_r, 829.abc\_s         & 17.4 & 40.4 & 8.2  & 41.2 & 1.8  & 15.2 & 1.7  & 3.7  & 20.0 & 19.7 & 7.8  & 18.4 & 1.9  & 6.7  & 453  & 3    \\
& 734.vpr\_r, 834.vpr\_s         & 17.2 & 21.8 & 9.4  & 10.3 & 2.4  & 2.4  & 1.9  & 2.0  & 29.6 & 16.9 & 12.0 & 10.0 & 3.6  & 1.7  & 662  & 2    \\
& 735.gem5\_r, 835.gem5\_s        & 22.9 & 18.3 & 2.4  & 1.7  & 0.1  & 0.0  & 0.1  & 0.0  & 35.3 & 17.7 & 4.3  & 1.9  & 0.6  & 0.2  & 134  & 0    \\
& 750.sealcrypto\_r   & 7.2  & --   & 1.9  & --   & 0.1  & --   & 0.1  & --   & 7.3  & --   & 2.3  & --   & 1.0  & --   & 496  & --   \\
& 753.ns3\_r, 853.ns3\_s         & 14.5 & 65.4 & 2.2  & 22.7 & 0.0  & 0.1  & 0.0  & 0.1  & 20.7 & 49.6 & 3.9  & 22.3 & 0.3  & 0.2  & 84   & 0    \\
& 777.zstd\_r         & 11.2 & --   & 8.0  & --   & 0.6  & --   & 0.6  & --   & 12.4 & --   & 6.7  & --   & 2.1  & --   & 491  & --   \\

\midrule

\multirow{12}{*}{\rotatebox[origin=c]{90}{\textbf{Floating Point}}}
& 709.cactus\_r, 809.cactus\_s       & 119.7& 133.4& 7.1  & 12.5 & 2.6  & 3.1  & 2.0  & 2.0  & 136.9& 104.8& 15.6 & 34.0 & 4.4  & 3.6  & 689  & 62   \\
& 722.palm\_r, 822.palm\_s         & 45.8 & 99.5 & 9.2  & 18.4 & 5.4  & 7.6  & 6.6  & 8.7  & 54.2 & 15.2 & 9.0  & 8.3  & 5.2  & 8.6  & 902  & 131  \\
& 731.astcenc\_r       & 5.0  & --   & 1.0  & --   & 0.0  & --   & 0.0  & --   & 6.4  & --   & 2.5  & --   & 0.1  & --   & 23   & --   \\
& 736.ocio\_r          & 6.2  & --   & 2.3  & --   & 2.3  & --   & 4.1  & --   & 15.3 & --   & 6.4  & --   & 1.5  & --   & 545  & --   \\
& 737.gmsh\_r          & 20.5 & --   & 14.9 & --   & 1.7  & --   & 1.0  & --   & 21.6 & --   & 9.2  & --   & 3.1  & --   & 494  & --   \\
& 748.flightdm\_r      & 32.3 & --   & 0.2  & --   & 0.0  & --   & 0.0  & --   & 41.7 & --   & 0.6  & --   & 0.0  & --   & 0    & --   \\
& 749.fotonik3d\_r, 849.fotonik3d\_s    & 63.5 & 31.7 & 48.2 & 31.1 & 44.2 & 20.1 & 41.1 & 25.4 & 58.9 & 26.7 & 45.8 & 13.4 & 24.1 & 14.0 & 885  & 508  \\
& 765.roms\_r, 865.roms\_s         & 198.3& 223.4& 116.7& 120.8& 27.2 & 59.1 & 14.0 & 24.4 & 157.5& 136.2& 102.9& 88.3 & 34.3 & 19.7 & 931  & 378  \\
& 766.femflow\_r       & 42.3 & --   & 7.7  & --   & 0.7  & --   & 0.9  & --   & 58.8 & --   & 12.7 & --   & 4.0  & --   & 879  & --   \\
& 767.nest\_r, 867.nest\_s         & 36.9 & 25.6 & 4.6  & 27.6 & 0.2  & 12.3 & 0.3  & 4.0  & 37.5 & 20.9 & 6.8  & 5.3  & 1.4  & 2.3  & 393  & 176  \\
& 772.marian\_r, 872.marian\_s       & 52.3 & 51.7 & 15.4 & 38.0 & 0.4  & 2.7  & 0.6  & 2.8  & 52.8 & 43.3 & 14.3 & 25.9 & 4.2  & 3.4  & 858  & 51   \\
& 782.lbm\_r           & 55.2 & --   & 28.8 & --   & 17.8 & --   & 22.8 & --   & 49.6 & --   & 29.0 & --   & 18.4 & --   & 870  & --   \\

\midrule

\multirow{5}{*}{\rotatebox[origin=c]{90}{\textbf{Integer}}}
& 801.xz\_s           & --   & 9.5  & --   & 9.5  & --   & 1.8  & --   & 0.9  & --   & 11.3 & --   & 9.5  & --   & 2.3  & --   & 28   \\
& 817.flac\_s         & --   & 1.3  & --   & 0.0  & --   & 0.0  & --   & 0.0  & --   & 3.4  & --   & 1.1  & --   & 0.1  & --   & 0    \\
& 838.diamond\_s      & --   & 0.4  & --   & 0.1  & --   & 0.0  & --   & 0.1  & --   & 3.2  & --   & 1.9  & --   & 0.1  & --   & 7    \\
& 846.minizinc\_s     & --   & 23.9 & --   & 6.0  & --   & 0.1  & --   & 0.1  & --   & 23.2 & --   & 7.3  & --   & 2.7  & --   & 0    \\
& 854.graph500\_s     & --   & 65.5 & --   & 59.5 & --   & 14.2 & --   & 10.6 & --   & 37.8 & --   & 27.3 & --   & 3.1  & --   & 94   \\

\midrule

\multirow{7}{*}{\rotatebox[origin=c]{90}{\textbf{Floating Point}}}
& 800.pot3d\_s         & --   & 263.3& --   & 180.8& --   & 230.6& --   & 47.8 & --   & 20.2 & --   & 14.7 & --   & 9.7  & --   & 30   \\
& 803.sph\_exa\_s      & --   & 50.5 & --   & 2.5  & --   & 1.1  & --   & 0.7  & --   & 48.6 & --   & 5.3  & --   & 1.5  & --   & 124  \\
& 811.tealeaf\_s       & --   & 22.3 & --   & 17.0 & --   & 16.0 & --   & 29.9 & --   & 13.2 & --   & 5.6  & --   & 1.1  & --   & 1    \\
& 816.nab\_s           & --   & 9.6  & --   & 0.8  & --   & 0.6  & --   & 0.4  & --   & 8.8  & --   & 1.8  & --   & 0.6  & --   & 31   \\
& 820.cloverleaf\_s    & --   & 161.0& --   & 63.1 & --   & 39.3 & --   & 37.1 & --   & 45.0 & --   & 37.5 & --   & 20.8 & --   & 557  \\
& 857.namd\_s          & --   & 11.2 & --   & 0.7  & --   & 0.6  & --   & 0.7  & --   & 32.4 & --   & 1.4  & --   & 0.9  & --   & 87   \\
& 881.neutron\_s       & --   & 60.0 & --   & 58.0 & --   & 10.5 & --   & 4.0  & --   & 30.9 & --   & 30.3 & --   & 3.7  & --   & 108  \\

\bottomrule
\end{tabular}

\vspace{2pt}
{\scriptsize \textit{Note:} MPKI = Misses per Kilo-Instructions. Mem. BW columns show memory bandwidth in GB/s (read + write). Empty cells indicate no corresponding rate/speed variant.}
\end{table*}

Across both rate and speed suites, a small set of benchmarks dominate cache pressure and downstream memory bandwidth demand. High L1D MPKI does not uniformly translate to high L3 MPKI, revealing substantial variation in cache filtering effectiveness. Workloads such as 709/809.cactus and 772/872.marian exhibit strong L2/L3 locality, while 765/865.roms, 749/849.fotonik3d, 782.lbm\_r, and 800.pot3d\_s generate sustained L3 misses, consistent with their backend-memory-bound behavior (Section~\ref{tmam}). Table \ref{tab:cache_filter_summary} summarizes benchmarks exhibiting the most pronounced cache filtering behavior.

In rate mode, poor cache filtering directly amplifies memory bandwidth demand as copy count scales. The highest 512-copy consumers: 765.roms\_r, 722.palm\_r, 749.fotonik3d\_r, 766.femflow\_r, and 782.lbm\_r, reach 870-931~GB/s, approaching the system theoretical bandwidth limit of approximately 614~GB/s per socket (1228~GB/s dual-socket)~\cite{amd2025epycwhitepaper}. This saturation explains the backend-memory bottleneck and retiring collapse observed in Section~\ref{tmam}. Notably, all workloads exceeding 850~GB/s at scale are floating-point benchmarks. Table \ref{tab:mem_bw_summary} highlights benchmarks with notable memory bandwidth scaling behavior.

Integer workloads benefit from effective cache filtering and rarely saturate memory bandwidth, resulting in more balanced TMA profiles at scale. Speed benchmarks exhibit a different scaling regime due to their shared-memory execution model: cache and coherence reuse reduce redundant DRAM traffic, and some workloads draw lower bandwidth at 512 threads than at 1 thread (e.g., 800.pot3d\_s). Consequently, even the highest speed-mode consumers remain well below their rate counterparts; for example, 865.roms\_s reaches 378~GB/s at 512 threads versus 931~GB/s at 512 copies for 765.roms\_r, a 2.5x difference driven by reduced redundancy and differing problem structure.

\begin{table}[tb]
\centering
\caption{Cache Filtering Effectiveness Summary}
\label{tab:cache_filter_summary}

\begin{tblr}{
  colspec = {l r X},
  width = \linewidth,
  row{1} = {font=\bfseries},
  hlines,
  vlines,
}
Benchmark & Reduction & Key Behavior \\
765.roms\_r      & 7.3$\times$   & Large working set; sustained DRAM traffic \\
709.cactus\_r    & 46$\times$    & Strong spatial and temporal locality \\
772.marian\_r    & 130$\times$   & Excellent L3 reuse (ML attention) \\
749.fotonik3d\_r & 1.4$\times$   & Streaming; low reuse \\
782.lbm\_r       & 3.1$\times$   & Irregular access; memory-bound \\
800.pot3d\_s     & 1.1$\times$   & Near-streaming; 87\% L3 miss rate \\
865.roms\_s      & 3.8$\times$    & Scaled-up rate behavior \\
809.cactus\_s    & 43$\times$    & Cache-efficient despite high L1 pressure \\
\end{tblr}

\vspace{2pt}
{\scriptsize Reduction = L1D MPKI / L3 MPKI.}
\end{table}

\begin{table}[tb]
\centering
\caption{Memory Bandwidth Scaling Summary}
\label{tab:mem_bw_summary}

\begin{tblr}{
  colspec = {l r X},
  width   = \linewidth,
  row{1}  = {font=\bfseries},
  hlines,
  vlines,
}
Benchmark & Scaling & Key Behavior \\

765.roms\_r       & 66.5$\times$  & Near socket bandwidth limit; backend memory-bound \\
722.palm\_r       & 136.7$\times$ & Large amplification with copy scaling \\
749.fotonik3d\_r  & 21.6$\times$  & Sustained high bandwidth from streaming access \\
766.femflow\_r    & 976.7$\times$ & Explosive scaling from minimal single-copy bandwidth \\
782.lbm\_r        & 37.8$\times$  & High DRAM demand consistent with memory-bound behavior \\

800.pot3d\_s      & 0.6$\times$   & Bandwidth decreases due to cache reuse \\
820.cloverleaf\_s & 15$\times$   & Highest speed mode bandwidth; stencil/grid workload \\
849.fotonik3d\_s  & 20$\times$   & Streaming HPC kernel scales with threads \\
865.roms\_s       & 15.5$\times$   & Substantially lower bandwidth than rate variant \\

\end{tblr}

\vspace{2pt}
{\scriptsize Scaling denotes the ratio of bandwidth at 512 copies/threads to bandwidth at 1 copy/thread.}
\end{table}

\subsection{Instruction Mix and SIMD Utilization}
\label{simd}

Instruction mix reveals the computational character of workloads and indicates whether compilers successfully autovectorize hot loops to exploit "Zen 5's" native 512-bit AVX-512 pipelines. Tables \ref{tab:simd_rate} and \ref{tab:simd_speed} present SIMD utilization for all SPEC CPU\textregistered2026 benchmarks along with their computational intensity (FLOPs).

\begin{table}[tbp]
\centering
\caption{SIMD Utilization \\ SPECrate\textregistered2026 Suites}
\label{tab:simd_rate}

\begin{tabular}{@{}c l rrrr r@{}}
\toprule
& \textbf{Benchmark} & \textbf{512} & \textbf{256} & \textbf{128} & \textbf{Sc.} & \textbf{GFLOPs} \\
\midrule
\multirow{14}{*}{\rotatebox[origin=c]{90}{\textbf{Integer}}}
& 706.stockfish\_r   & 39.8 & 15.3 & 44.9 & 0.0  & 0.00 \\
& 707.ntest\_r       & 6.2  & 15.8 & 77.0 & 1.0  & 0.00 \\
& 708.sqlite\_r      & 4.9  & 19.2 & 67.7 & 8.2  & 0.03 \\
& 710.omnetpp\_r     & 2.1  & 32.0 & 52.0 & 13.9 & 0.06 \\ 
& 714.cpython\_r     & 8.3  & 5.6  & 84.1 & 2.1  & 0.00 \\
& 721.gcc\_r         & 11.0 & 5.9  & 83.0 & 0.1  & 0.00 \\
& 723.llvm\_r        & 6.6  & 11.5 & 81.9 & 0.0  & 0.00 \\
& 727.cppcheck\_r    & 0.9  & 53.0 & 46.0 & 0.1  & 0.00 \\
& 729.abc\_r         & 5.0  & 16.3 & 67.9 & 10.8 & 0.03 \\
& 734.vpr\_r         & 10.0 & 9.5  & 62.8 & 17.7 & 0.26 \\
& 735.gem5\_r        & 7.6  & 9.3  & 69.3 & 13.8 & 0.04 \\
& 750.sealcrypto\_r  & 61.6 & 0.0  & 38.4 & 0.0  & 0.00 \\
& 753.ns3\_r         & 3.6  & 11.2 & 63.2 & 22.0 & 0.21 \\
& 777.zstd\_r        & 2.0  & 0.0  & 97.5 & 0.5  & 0.00 \\
\midrule
\multirow{12}{*}{\rotatebox[origin=c]{90}{\textbf{Floating Point}}}
& 709.cactus\_r      & 0.4  & 0.1  & 36.4 & 63.2 & 6.22 \\
& 722.palm\_r        & 8.8  & 3.4  & 51.8 & 36.0 & 7.86 \\
& 731.astcenc\_r     & 0.3  & 0.5  & 54.7 & 44.6 & 4.56 \\
& 736.ocio\_r        & 3.5  & 0.0  & 49.7 & 46.8 & 5.86 \\
& 737.gmsh\_r        & 0.8  & 3.8  & 64.3 & 31.0 & 1.04 \\
& 748.flightdm\_r    & 1.1  & 5.7  & 56.9 & 36.4 & 1.46 \\
& 749.fotonik3d\_r   & 11.5 & 8.9  & 73.0 & 6.7  & 9.49 \\
& 765.roms\_r        & 70.5 & 2.7  & 18.4 & 8.4  & 10.37 \\
& 766.femflow\_r     & 41.1 & 4.1  & 28.3 & 26.6 & 23.42 \\
& 767.nest\_r        & 0.5  & 1.6  & 53.3 & 44.7 & 2.45 \\
& 772.marian\_r      & 15.3 & 3.8  & 46.3 & 34.7 & 2.18 \\
& 782.lbm\_r         & 0.0  & 0.0  & 28.8 & 71.2 & 8.58 \\
\bottomrule
\end{tabular}

\vspace{2pt}
{\scriptsize 512/256/128 = AVX-512/256/128 (\%), Sc = Scalar (\%) and GFLOPs for 1-copy runs.}
\end{table}

\begin{table}[tbp]
\centering
\caption{SIMD Utilization \\ SPECspeed\textregistered2026 Suites}
\label{tab:simd_speed}
\begin{tabular}{@{}c l rrrr r@{}}
\toprule
& \textbf{Benchmark} & \textbf{512} & \textbf{256} & \textbf{128} & \textbf{Sc.} & \textbf{GFLOPs} \\
\midrule
\multirow{13}{*}{\rotatebox[origin=c]{90}{\textbf{Integer}}}
& 801.xz\_s          & 8.2  & 2.0  & 89.8 & 0.0  & 0.00 \\
& 807.ntest\_s       & 1.8  & 40.1 & 56.6 & 1.5  & 0.00 \\
& 817.flac\_s        & 69.9 & 5.0  & 19.5 & 5.7  & 0.47 \\
& 821.gcc\_s         & 10.7 & 6.0  & 83.2 & 0.1  & 0.00 \\
& 823.llvm\_s        & 9.3  & 18.2 & 72.5 & 0.0  & 0.00 \\
& 827.cppcheck\_s    & 0.5  & 45.1 & 54.1 & 0.3  & 0.00 \\
& 829.abc\_s         & 10.5 & 48.1 & 41.3 & 0.2  & 0.00 \\
& 834.vpr\_s         & 10.6 & 5.6  & 66.1 & 17.8 & 0.29 \\
& 835.gem5\_s        & 23.1 & 8.3  & 58.1 & 10.6 & 0.02 \\
& 838.diamond\_s     & 16.9 & 1.3  & 77.2 & 4.6  & 0.03 \\
& 846.minizinc\_s    & 16.2 & 9.1  & 72.0 & 2.7  & 0.00 \\
& 853.ns3\_s         & 12.2 & 12.5 & 59.1 & 16.2 & 0.14 \\
& 854.graph500\_s    & 22.4 & 8.8  & 68.7 & 0.0  & 0.00 \\
\midrule
\multirow{13}{*}{\rotatebox[origin=c]{90}{\textbf{Floating Point}}}
& 800.pot3d\_s       & 54.1 & 11.8 & 9.7  & 24.4 & 4.73 \\
& 803.sph\_exa\_s    & 3.8  & 0.4  & 22.0 & 73.8 & 5.54 \\
& 809.cactus\_s      & 0.6  & 0.0  & 36.6 & 62.7 & 5.63 \\
& 811.tealeaf\_s     & 0.0  & 0.0  & 28.2 & 71.9 & 6.48 \\
& 816.nab\_s         & 0.0  & 0.0  & 37.8 & 62.2 & 4.11 \\
& 820.cloverleaf\_s  & 24.2 & 0.0  & 30.0 & 45.8 & 7.23 \\
& 822.palm\_s        & 10.7 & 2.5  & 53.7 & 33.0 & 8.37 \\
& 849.fotonik3d\_s   & 0.3  & 6.4  & 49.7 & 43.6 & 6.69 \\
& 857.namd\_s        & 0.0  & 0.0  & 26.3 & 73.7 & 9.74 \\
& 865.roms\_s        & 65.7 & 2.6  & 21.5 & 10.2 & 8.32 \\
& 867.nest\_s        & 0.5  & 1.7  & 52.3 & 45.5 & 0.69 \\
& 872.marian\_s      & 16.6 & 4.7  & 57.9 & 20.8 & 0.94 \\
& 881.neutron\_s     & 0.1  & 4.0  & 45.3 & 50.6 & 1.88 \\
\bottomrule
\end{tabular}

\vspace{2pt}
{\scriptsize 512/256/128 = AVX-512/256/128 (\%), Sc = Scalar (\%) and GFLOPs for 1-thread runs.}
\end{table}

The highest AVX-512 utilization among FP benchmarks are 765.roms\_r, 766.femflow\_r, 772.marian\_r, and 749.fotonik3d\_r. The 765.roms\_r benchmark (ocean modeling) demonstrates excellent vectorization of stencil computations with 10.4 GFLOPs throughput.

Among integer benchmarks, 750.sealcrypto\_r achieves 61.6\% AVX-512 utilization for wide-integer arithmetic in homomorphic encryption. Notably, we find that 706.stockfish\_r (chess engine) achieves 39.8\% AVX-512 utilization for SIMD bitboard operations, demonstrating that AVX-512 benefits extend beyond traditional floating-point HPC workloads to algorithmic integer computations.

The most scalar-dominated FP workloads are 782.lbm\_r, 709.cactus\_r, 736.ocio\_r, and 767.nest\_r. Despite being FP-intensive, these workloads operate predominantly in scalar or 128-bit mode.

Speed integer benchmarks include several standout vectorization cases. The highest AVX-512 utilization is 817.flac\_s demonstrating autovectorization success for audio encoding signal processing. Speed FP benchmarks show 865.roms\_s and 800.pot3d\_s consistent with the rate variant behavior and reflecting successful compiler vectorization of stencil computations.

Interestingly, several speed integer benchmarks show substantial AVX-256 utilization: 829.abc\_s, 827.cppcheck\_s and 807.ntest\_s suggesting wider vector operations for string/pattern matching and analysis tasks.

\textbf{FLOPs Distribution.} The highest floating-point throughput workloads are 766.femflow\_r, 765.roms\_r, 749.fotonik3d\_r, 782.lbm\_r, and 722.palm\_r. The 766.femflow\_r benchmark achieves the highest GFLOPs with 41.1\% AVX-512 utilization. Notably, 782.lbm\_r achieves 8.6 GFLOPs with 0\% AVX-512, operating entirely in scalar/128-bit mode, indicating a vectorization opportunity not exploited by the compiler.

While FP benchmarks naturally dominate FLOPs throughput, integer benchmarks show substantial SIMD utilization across all vector widths. The highest AVX-512 users are 750.sealcrypto\_r and 706.stockfish\_r, while 727.cppcheck\_r favors AVX-256 and 710.omnetpp\_r uses a mix of widths. This demonstrates that modern integer workloads actively leverage SIMD for operations like string matching, bitwise operations, and memory operations.

\section{Synthesis and Design Implications}
\label{synthesis}

The four characterization lenses of Section \ref{characterization} - TMA bottleneck distribution, control flow characteristics, cache hierarchy behavior, and SIMD utilization - each illuminate distinct aspects of benchmark behavior. Individually, these metrics identify bottlenecks; combined, they reveal why workloads behave as they do and what design trade-offs govern their performance. Using the four lenses, this section groups the workloads into behavioral clusters and observes the performance of the "Zen 5" microarchitecture based on its design trade-offs.

\subsection{Workload Clustering and Taxonomy}
\label{clustering}

\textbf{Frontend Control-Flow-Dominated Integer.} \textit{727.cppcheck\_r, 753.ns3\_r, 714.cpython\_r, 735.gem5\_r, 854.graph500\_s} - These workloads combine high Frontend Bound (Section \ref{tmam}) with exceptionally high branch density (Section \ref{controlflow}) yet maintain low misprediction rates. The key insight is that they stress branch predictor throughput rather than accuracy.

\textbf{High-Efficiency Compute.} \textit{750.sealcrypto\_r, 736.ocio\_r, 707.ntest\_r, 767.nest\_r, 766.femflow\_r} - These achieve the highest Retiring at 1-copy (Section \ref{tmam}), demonstrating effective utilization of "Zen 5's" wide dispatch. At 512-copy, these workloads exhibit the highest SMT resource sharing, as their efficiency at filling dispatch slots leaves minimal idle bandwidth for thread interleaving. Notably, this cluster includes both vector-intensive workloads (750.sealcrypto\_r, 766.femflow\_r) and scalar-dominated ones (767.nest\_r, 736.ocio\_r), demonstrating that dispatch efficiency, not vector width, is the unifying characteristic.

\textbf{Memory BW Bound FP.} \textit{765/865.roms, 749/849.fotonik3d, 722.palm\_r, 782.lbm\_r} - These dominate Backend Memory at 512-copy (and speed) (Section \ref{tmam}) and saturate memory bandwidth (Section \ref{cache&mem}). These benchmark's performance scales directly with memory subsystem throughput.

\textbf{772.marian\_r Case Study.} We find that the neural machine translation workload exhibits dramatic copy-count-dependent behavior. At 1-copy, 772.marian\_r achieves 43.0\% Retiring (Section \ref{tmam}). At 512-copy, Retiring collapses to 13.2\% while L3 MPKI increases 10.3x. The shift to Backend-memory (51.4\%), despite moderate bandwidth consumption (Section \ref{cache&mem}), indicates L3 cache contention, as 512 independent inference instances compete for the shared 32MB L3 per CCD, evicting each other's working sets. This highlights a critical consideration for server-scale ML inference deployment.

\subsection{"Zen 5" Specific Observations}
\label{zen5obs}

\textbf{SMT Resource Sharing.} Correlating TMA with SIMD data reveals that some of the highest vector utilization benchmarks exhibit the highest SMT resource utilization. These benchmarks saturate "Zen 5's" 8-wide dispatch leaving no room for thread interleaving.

\textbf{Native AVX-512 Efficiency.} High AVX-512 utilization correlates with high Retiring, confirming native 512-bit execution without throttling. The 765.roms\_r exception (highest AVX-512, low Retiring) demonstrates that memory bandwidth, not instruction supply, becomes the constraint for streaming workloads.

\textbf{L3 Topology Effects.} The 772.marian\_r (capacity interference, Section~\ref{clustering}) versus 765.roms\_r (bandwidth saturation) contrast shows that "memory-bound" workloads require different remedies: L3 partitioning versus memory subsystem improvements.

\subsection{Implications for Architects and Software Developers}
\label{implications}

The characterization data surfaces actionable directions beyond the ``Zen~5''-specific observations in Section~\ref{zen5obs}.

\textbf{For Architects.} The architecturally salient result is less any single cluster's bottleneck than its \textit{scale-dependence}. The effects that most shape full-system behavior, SMT dispatch contention among the most dispatch-efficient workloads (19.3\% average throughput reduction at 512-copy) and shared-L3 capacity interference, are absent at single-copy and emerge only at full utilization. Not every memory bottleneck is scale-emergent, however: the Memory BW Bound cluster already exhibits poor L3 filtering at single-copy (fotonik3d: 1.4$\times$, lbm: 3.1$\times$ - Table~\ref{tab:cache_filter_summary}), reflecting inherent low-reuse streaming rather than contention. Separating these scale-emergent effects from structural ones is a prerequisite for design trade-off evaluation that single-instance characterization would otherwise misweight.

\textbf{For Software Developers.} Two practical recommendations follow from the data. For ML inference on EPYC, the 772.marian\_r case study (Section~\ref{clustering}) shows that keeping inference instance working sets within a single CCD's 32\,MB L3 avoids the pipeline inefficiency observed when 512 concurrent instances compete for shared L3 capacity, causing Retiring to drop from 43.0\% to 13.2\%. More broadly, our GCC-based results represent one point in the compiler design space rather than an upper bound; toolchain choice is itself a performance variable for ``Zen'' microarchitectures, particularly for HPC floating-point and compute-dense integer workloads. Comparing vendor-tuned toolchains such as the AMD Optimizing C/C++ and Fortran Compilers (AOCC) against this GCC baseline is therefore a natural extension of this work.

\section{Conclusion}
\label{conclusion}

This paper presents the first microarchitecture based performance characterization of SPEC CPU\textregistered2026 on AMD EPYC\texttrademark\space "Zen 5", offering two primary contributions: the first such characterization on any platform, and a scale analysis methodology comparing single-instance to full-system behavior that exposes bottlenecks invisible to conventional characterization. Using a multi-lens methodology spanning pipeline efficiency, control flow behavior, cache and memory subsystem pressure, and instruction mix, we analyze both SPECrate\textregistered (at single-copy and 512-copy scales) and SPECspeed\textregistered suites.

SPEC CPU\textregistered2026 exhibits substantial behavioral diversity, with IPC spanning a 3.3x range across benchmarks. Multi-lens analysis reveals distinct behavioral clusters: frontend control-flow-dominated integer workloads stress branch predictor throughput rather than accuracy; high-efficiency compute workloads achieve exceptional dispatch utilization but suffer corresponding SMT contention at scale; and memory bandwidth-bound floating-point workloads exhaust DRAM bandwidth at scale while exhibiting low L3 hit rates even at single-copy. The 772.marian benchmark exemplifies scale-dependent behavior: L3 capacity interference causes Retiring to collapse while bandwidth remains moderate, a distinction from bandwidth saturation that has direct implications for multi-tenant ML inference deployment. These effects emerge only at scale and highlight the importance of system-level characterization.

Scale analysis proves essential for understanding server-class processor behavior. At 512-copy, we observe SMT dispatch contention averaging 19\% reduction in effective throughput for benchmarks that most efficiently utilize "Zen 5's" wide dispatch. Parallelism analysis reveals fundamental divergence in scaling behavior across the SPECspeed\textregistered suite, from near-perfect thread utilization in several benchmarks to essentially serial execution in others, a diversity that reflects the breadth of application domains now represented in SPEC CPU\textregistered2026.

This characterization establishes a foundation for several promising research directions. Future studies should explore power efficiency across the benchmark suite, particularly given the distinct microarchitectural behaviors observed between integer and floating-point workloads. The cache contention patterns we identified motivate noisy-neighbor analyses essential for understanding VM isolation and cloud multi-tenancy. Building on the developer guidance in Section~\ref{implications}, systematic compiler sensitivity studies comparing GCC with vendor-optimized toolchains such as the AMD Optimizing C/C++ and Fortran Compilers (AOCC) would clarify how toolchain choice affects the behaviors we characterize. Finally, memory technology and sensitivity studies merit investigation as these directly impact both bandwidth-sensitive benchmarks and datacenter power budgets. By pursuing these directions, researchers can build on the architectural baseline established here to inform the design of next-generation datacenter processors.

\section*{Acknowledgment}
The authors thank their managers at AMD - Joseph Jensen, Josh Rovang, and Anil Saproo - for their support, which made this work possible. We further thank Milind Damle for his thorough review of this paper, and David Reiner for his guidance in ensuring this work adheres to SPEC's fair-use and trademark guidelines.

\bibliographystyle{IEEEtranS}
\bibliography{reference}

\vspace{6pt}
\noindent\rule{\linewidth}{0.4pt}
{\scriptsize \noindent SPEC\textregistered, SPEC CPU\textregistered,
SPECrate\textregistered, and SPECspeed\textregistered\space are registered trademarks of the Standard Performance Evaluation Corporation. See \url{https://www.spec.org} for more information about SPEC\textregistered\space benchmarks.}

\end{document}